\documentclass[a4paper,12pt]{article}
\usepackage{amsmath,amssymb,amsfonts,amsthm}
\newtheorem{theorem}{Theorem}[section]

\newtheorem{corollary}[theorem]{Corollary}

\usepackage{t1enc}
\usepackage[english]{babel}
\usepackage[latin1]{inputenc}

\title{Bounds on the force between black holes}
\author{Mar\'ia E. Gabach Clement\\
Max Planck Institute for Gravitational Physics,\\ (Albert Einstein Institute), Am M\"uhlenberg 1,\\ D-14476 Golm, Germany.}

\begin{document}

\maketitle

\begin{abstract}
We treat the problem of $N$ interacting, axisymmetric black holes and obtain two relations among physical parameters of the system including the force between the black holes. The first relation involves the total mass, the angular momenta, the distances and the forces between the black holes. The second one relates the angular momentum and area of each black hole with the forces acting on it. 
\end{abstract}
\section{Introduction}

The problem of interacting black holes dates back to the beginnings of General Relativity. Although a solution representing multiple black holes in equilibrium is known to exist \cite{ChruscielGalloway10} (the Majumdar-Papapetrou solution, which consists of $N$ extreme Reissner-N\"ordstrom black holes, i.e. each black hole has electric charge equal to its mass parameter, $q_i=m_i$), it is in general expected that there should be a force between different components of the horizon in order to prevent the spacetime from collapsing. Nevertheless, based on Newtonian ideas, a main concern was the following: Given two black holes, could the spin-spin repulsion compensate for the gravitational attraction and keep the system in equilibrium? Many attempts have since then been made in order to answer this question (see \cite{Bach21}, \cite{Weinstein90}, \cite{Li91} \cite{Varzugin97}, \cite{DainOrtiz09}, \cite{Manko08}). In particular, Li and Tian \cite{Li91} and Neugebauer and Hennig \cite{Neugebauer11}, \cite{Neugebauer11b} \cite{Neugebauer09} proved that an axially symmetric, stationary, vacuum solution of Einstein equation with disconnected horizon (and only two connected components) can not exist in equilibrium. Weinstein \cite{Weinstein90}, \cite{Weinstein92}, \cite{Weinstein94}, using harmonic maps, showed that asymptotically flat, vacuum, axially symmetric multi black hole solutions exist, although possibly with a conical singularity in the bounded component of the symmetry axis. 

The appearence of this conical singularity on the axis was clearly seen in the superposition of two Schwarzschild black holes in the early work of Bach \cite{Bach21} and was later discussed by Einstein and Rosen in \cite{Einstein36}. Although the existence of a singularity may seem discouraging, the solution so built has a well defined action and the corresponding thermodynamical potential describes a system of black holes at fixed temperature and at a fixed distance \cite{Herdeiro09}, \cite{Costa00}. The long range interactions in the multiple black hole solutions are unable to provide equilibrium between the various black holes and the conical singularity is interpreted as a boundary condition that keep the holes at fixed separation preventing the collapse of the system.

The non-existence proof of Neugebauer and Hennig \cite{Neugebauer11} for two rotating black holes in vacuum strongly uses the relation $A\geq8\pi|J|$ between the area and the angular momentum of each black hole  which is known to hold in the axisymmetric, regular case (see \cite{Acena11}). This type of geometrical inequalities relating physical parameters  of black holes have been extensively studied in the last few years as an attempt to explore the allowed values of angular momentum, mass and size a black hole can have,  and turned out to be relevant in different contexts as well (see \cite{Dainreview11} for an excellent recent review on geometrical inequalities). As interesting outcomes of these researches, two main relations have been found.

The first one of such relations in the context of multiple black holes configurations was proved by Chrusciel \textit{et al} \cite{Chrusciel08} (see also \cite{Dain06} for the inequality in the case of $N=1$),  and gives a lower bound to the ADM mass in the axisymmetric, vacuum case in the form
\begin{equation}\label{mj2}
0\geq f(J_1,..,J_N)-m
\end{equation}  
where $m$ is the total (ADM) mass, $J_i$ are the Komar angular momenta of the individual constituents and $f$ is a function of the $J_i$ (and possibly depending on other parameters as well), not known explicitely, which reduces in the case of a single black hole to $\sqrt{|J|}$.

The second inequality we are interested in, and the one used in \cite{Neugebauer11} is the relation
\begin{equation}\label{aj1}
1\geq \frac{8\pi |J|}{A},
\end{equation}
between the angular momentum $J$ and the area $A$ of an axisymmetric black hole which has recently been proven. In \cite{Acena11}, \cite{Gabach11} and \cite{Dain-Reiris11} maximal vacuum initial data (possibly with a positive cosmological constant) containing a minimal stable surface are considered. Also, in \cite{Jaramillo11} the inequality is proven for spacetimes having outermost stably marginally trapped surfaces, not necessarily in vacuum. As we will discuss later, \eqref{aj1} has been extended to incorporate electric and magnetic charges. Inequality \eqref{aj1} holds for each black hole in a regular initial data and, loosely speaking, states that a hole can not rotate too fast. 

An important ingredient in the above results, relations \eqref{mj2} and \eqref{aj1}, is the assumed regularity at the symmetry axis. As we mentioned, this is related to the notion of equilibrium. We will incorporate the interaction between the black holes into the problem is by means of relaxing the regularity condition at the axis and therefore allowing a conical singularity  between the holes. The value of the deficiency angle will be related to the gravitational force \cite{Weinstein90}. The purpose of this work is to investigate how inequalities \eqref{mj2} and \eqref{aj1} are modified when one takes into account explicitely the interaction between the black holes, that is, when one includes non-equilibrium configurations. In this way we will obtain bounds on the forces between the black holes (i.e. on the deficit angles) set by the physical quantities $A, m,$ and $J$.

The article is organized as follows. In section \ref{main} we briefly introduce the basic features of multiple black hole solutions to Einstein constraint equations and the explicit relation between force and deficit angle. Then, in section \ref{mass} we present the more especific hypothesis needed in order to obtain a relation between mass, angular momentum and force, similar in espirit to \eqref{mj2}, theorem \ref{teomass} and finally, in section \ref{area} we find a new inequality relating area, angular momentum and force, theorem \ref{teoarea}, analogous to \eqref{aj1}. 

\section{Main result}\label{main}

We begin this section with a brief description of multiple black hole solutions of Einstein equations in the form of maximal axisymmetric initial data. 

Consider a maximal, simply connected initial data set ($M, \,g_{ab}\,K_{ab}$), invariant under a U(1) action. It is known \cite{Chrusciel08a} that there exists a global coordinate system ($\rho,z,\phi$) such that the metric takes the form
\begin{equation}\label{3metric}
g=e^{\sigma+2q}(d\rho^2+dz^2)+\rho^2e^\sigma(d\phi+\rho A_\rho d\rho+A_zdz)^2,
\end{equation}
where the functions are taken to be $\phi-$independent.

We assume vacuum for simplicity, although matter satisfying some positive energy condition would be also allowed. The constraint equations in the maximal case, $g^ {ab}K_{ab}=0$, are
\begin{equation}
^3R=K_{ab}K^ {ab},\qquad \nabla^ aK_{ab}=0. 
\end{equation}
Due to these constraints, there exists a twist potential $\omega$ related to the second fundamental form  by the expression (see \cite{Dain06})
\begin{equation}
\epsilon_{abc}K^b_e\eta^c\eta^edx^a=d\omega
\end{equation}
where $\eta^a$ is a $2\pi-$periodic Killing vector field and $\epsilon_{abc}$ is the Levi Civita tensor. Moreover, we have the useful bound \cite{Dain06}
\begin{equation}\label{boundk}
K_{ab}K^ {ab}\geq e^{-\sigma-2q}\frac{|\partial\omega|^2}{2\eta^2}.
\end{equation}

The quantities that we are interested in in this work are the total mass of the initial data, and the area and angular momentum of each black hole. As is well known, the mass is a global quantity, given by a boundary integral at spatial infinity
\begin{equation}
 m=\frac{1}{16\pi}\lim_{r\to\infty}\int_{\partial B_r}(\partial_b g_{ab}-\partial_ag_{bb})n^adS
\end{equation}
where $r=\sqrt{\rho^2+z^2}$ is the Euclidean distance, $\partial$ denotes partial derivatives, $B_r$ is the Euclidean ball $r=const$, $\partial B_r$ is its border, $n^a$ is its exterior unit normal and $dS$ is the unit surface element with respect to the Euclidean metric, $dS=\sin\theta d\theta d\phi$.

Now, consider an arbitrary 2-surface $\Sigma$ in $M$ with induced metric $\gamma_{ab}$, then the area of $\Sigma$ is given by the expression
\begin{equation}\label{areaf}
 A(\Sigma)=\int_\Sigma dS_\gamma,
\end{equation}
with $dS_\gamma$ the surface element with respect to $\gamma_{ab}$.

Finally, for axially symmetric and maximal data one can define the angular momentum $J$ associated with $\Sigma$ (the Komar integral of the Killing vector) as
\begin{equation}
 J(\Sigma)=\int_\Sigma K_{ab}\eta^an^bdS_\gamma,
\end{equation}
where $n^a$ is the unit normal vector to $\Sigma$.

For concreteness, in what follows, we consider $N\geq1$ black holes located at the symmetry axis, which will be represented by punctures in section \ref{mass} and by minimal stable surfaces in section \ref{area}. The axis $\rho=0$ is denoted by $\Gamma$. It contains $N-1$ bounded components denoted by $\Gamma_i$,  $i=1,..,N-1$, and two unbounded components, $\Gamma_0$ and $\Gamma_N$.

As we mentioned in the introduction, the interaction between the black holes can be realised through the non regular character of the solution at $\Gamma$. The regularity condition at $\Gamma$ can be stated as
\begin{equation}\label{reg1}
 \lim_{\rho_0\to0^ +}\frac{\sqrt{\eta}}{\int_0^{\rho_0}e^ {\sigma/2+q}d\rho}=1
\end{equation}
where $\eta=\eta^a\eta_a$, and is translated into the requirement that the metric function $q$ (see the expression \eqref{3metric}) vanishes at the axis: $q|_{\Gamma}=0$.

In general this condition will not be satisfied and the metric will not define a regular solution of Einstein equations. If \eqref{reg1} does not hold, then the singularity at the axis is called a conical singularity. The deficit angle at the axis is given by $e^ {-q_i}$ where $q_i$ is the constant value of the function $q$ at the $i-$th connected component of the axis. These singularities are interpreted as forces needed to balance the gravitational attraction and keep the bodies in equilibrium and can be expressed in terms of $q_i$ as (see \cite{Weinstein94} for more details)
\begin{equation}\label{force}
 \mathcal F_i=\frac{1}{4}(e^{-q_i}-1).
\end{equation}
One would expect this force to be positive, reflecting the fact that a positive force is needed in order to prevent the black holes from falling on each other. A positive force has been found in a few situations, like in the zero \cite{Bach21} and small angular momentum case \cite{Li92}, the case when the problem admits an involutive symmetry \cite{Li91}, when there are two equal, counter-rotating Kerr black holes \cite{Varzugin98} and in the limit when one of the black holes becomes extreme and the distance to one of the adjacent black holes tends to zero \cite{Weinstein94}.
Nevertheless, so far there is no general proof stating that the force is always positive.

One can prescribe the value of the function $q$ on one of the component of the axis \cite{Weinstein94}, the other values will be determined by the solution of the constraints, and are not known in general. In this article, we give two results, which are the extensions of inequalities \eqref{mj2} and \eqref{aj1} to include the non zero values of $q$ at the axis and are presented in the following sections.

\subsection{Bound in terms of mass and angular momentum}\label{mass}

In order to extend Chrusciel \textit{et al} result \cite{Chrusciel08}, inequality \eqref{mj2}, to the non-regular case, we state here their main assumptions. We will retain basically all of them, and only relax the regularity of the function $q$ at the axis. 

As we mentioned before, for the mass-angular momentum inequality we represent the black holes as $N$ punctures on the axis, separated by distances $L_i$, each puncture being an asymptotically flat end in the sense that for $k\geq6$,
\begin{equation}\label{af1}
 g_{ab}-\delta_{ab}=o_k(r^{-1/2}),
\end{equation}
\begin{equation}\label{af2}
\partial_kg_{ab}\in L^2(M)
\end{equation}
where $f=o_k(r^\lambda)$ means that $f$ satisfies
\begin{equation}
 \partial_{k_1}...\partial_{k_n}f=o(r^{\lambda-n}),\qquad 0\leq n\leq k.
\end{equation}
The main result of this section gives a relation among the total mass of the initial data, the individual angular momenta of the each asymptotic region and the forces between them.
\begin{theorem}\label{teomass}
Consider a maximal, asymptotically flat, axisymmetric initial data ($M, g_{ab}, K_{ab}$) as described above, with $N$ extra asymptotically flat ends represented by punctures on the axis $\Gamma$, with angular momenta $J_i$, $i=1,..,N$. Then, there is a lower bound to the ADM mass $m$ given by
\begin{equation}\label{ine1}
\sum_{i=1}^ {N-1}\frac{L_i}{4}\ln(1+4\mathcal F_i)\geq f(J_1,...,J_N)-m
\end{equation}
where $L_i$ is the Newtonian length of $\Gamma_i$, $\mathcal F_i$ is the force between adjacent punctures and $f$ is a function of the $J_i$'s,.
\end{theorem}

Before entering into the proof, we want to make a few remarks about the result. It is clearly seen from inequality \eqref{ine1} that it does not imply, in general, the result $0\geq f-m$. That is only the case when the forces are non positive. But if the forces are positive, which is the expected situation, in principle, the total mass $m$ could take values lower than $f$, as opposed to the regular case \cite{Chrusciel08}.

As the data are asymptotically flat, we must have $q_0=q_N=0$, and \eqref{ine1} reduces to Chrusciel \textit{et al} result, $m\geq f$, for $q_i=0$ and also to Dain's result, $m\geq\sqrt{|J|}$, for $N=1$ \cite{Dain06}.

Note that there is no rigidity statement in this theorem. This is due in part, to the lack of a known minimizer for a mass functional appearing in the proof of Chrusciel et al result in the case $N\geq2$. But also, because in inequality \eqref{ine1} there appear two quantities that depend on the given parameters of the data. Both, the mass and the force are to be computed after solving the constraints and can not be prescribed a priori. Therefore, for given values of angular momenta $J_i$ and separation distances $L_i$, there might exist different solutions saturating the inequality with different values of $\mathcal F_i$ and mass. In this respect, it is worth mentioning, that there exists an explicit solution describing two equal, counter-rotating extreme Kerr black holes given by Manko et al \cite{Manko08} which could saturate \eqref{ine1}. We will meet this solution again in the following section.

\begin{proof}

Since the proof is an adaptation of Chrusciel \textit{et al} argument, \cite{Chrusciel08}, we only sketch the main steps and highlight the point where the non zero force is incorporated. We refer the reader to \cite{Chrusciel08} and its references for more details.

Begin with the expression for the curvature scalar  $R$ of the metric $g_{ab}$, given by
\begin{equation}\label{ricci}
-\frac{1}{8}Re^{\sigma+2q}=\frac{1}{4}\Delta\sigma+\frac{1}{16}(\partial\sigma)^2+\frac{1}{4}\Delta_2q+\frac{1}{16}\rho^2e^{-2q}(\rho A_{\rho,z}-A_{z,\rho})^2,
\end{equation}
where $\Delta$ is the Laplacian in $\mathbb R^3$ and $\Delta_2$ is the 2-dimensional Laplacian
\begin{equation}
\Delta_2q=q_{,\rho\rho}+q_{,zz}.
\end{equation}
Integrate \eqref{ricci} over $\mathbb R^3$ and use the asymptotic flatness condition and the definition of the mass $m$ (see \cite{Dain06}):
\begin{equation}\label{masssigma}
 \int_{\mathbb R^3}\Delta\sigma d^3x=\lim_{r\to\infty}\int_{\partial B_r}\partial_r\sigma r^ 2dS=-8\pi m,
\end{equation}
to obtain
\begin{equation}
2\pi m=\int_{\mathbb R^3}\left[\frac{1}{16}(\partial\sigma)^2+\frac{1}{4}\Delta_2q +\frac{1}{8} Re^{\sigma+2q}+\frac{1}{16}\rho^2e^{-2q}(\rho A_{\rho,z}-A_{z,\rho})^2\right]d^3x.
\end{equation}
Use inequality \eqref{boundk} to bound the Ricci scalar in the above integral and disregard the last, non negative term to obtain the following lower bound for the mass
\begin{equation}\label{mass1}
m\geq\frac{1}{32\pi}\int_{\mathbb R^3}(\partial\sigma)^2+\frac{|\partial\omega|^2}{\eta^2}dx^3+\frac{1}{8\pi}\int_{\mathbb R^3}\Delta_2qd^3x.
\end{equation}
Note that the first integral is exactly the functional $\mathcal M$ used by Dain \cite{Dain06} and Chrusciel et al \cite{Chrusciel08} in the proof of the mass-angular momentum inequalities,
\begin{equation}
\mathcal M:=\frac{1}{32\pi}\int_{\mathbb R^3}(\partial\sigma)^2+\frac{(\partial\omega)^2}{\eta^2}dx^3.
\end{equation}
The second integral in the right hand side of \eqref{mass1} is zero for regular solutions (i.e, when $q=0$ at the axis). But when the solution contains conical singularities it gives instead
\begin{equation}
 \int_{\mathbb R^3}\Delta_2qd^3x=2\pi\sum_{i=1}^{N-1}q_iL_i
\end{equation}
where $L_i$ is the Newtonian length of $\Gamma_i$ and $q_i$ is the constant value of $q$ on it. With this result, we arrive at the inequality
\begin{equation}
m\geq\mathcal M+\frac{1}{4}\sum_{i=1}^{N-1}q_iL_i.
\end{equation}

In \cite{Chrusciel08} (see proposition 2.1) the authors prove that for any set of aligned punctures and of axis values $\omega_i:=\omega|_{\Gamma_i}$, there exists a solution ($\tilde\sigma, \tilde\omega$) of the variational equations associated with the functional $\mathcal M$, with finite values of $\mathcal M$ and appropriate asymptotic behavior near each puncture. 
So, denote by $f(J_1,..J_N)$ the numerical value of $M$ of the harmonic map, $(\tilde\sigma,\tilde\omega)$ from $\mathbb R^3\setminus\Gamma$ to the two dimensional hyperbolic space. Then we obtain
\begin{equation}\label{ecc}
m\geq f(J_1...J_N)+\frac{1}{4}\sum_{i=1}^ {N-1}q_iL_i.
\end{equation}
Finally expressing the constants $q_i$ in \eqref{ecc} in terms of the forces (see equation \eqref{force}), the statement of the theorem is proven. 
\end{proof}

\subsection{Bound in terms of area and angular momentum}\label{area}
In this section we obtain a bound on the forces acting on a black hole in terms of its area and angular momentum. This will generalize inequality \eqref{aj1} to the non-equilibrium case.

For that purpose, we will follow Dain and Reiris's argument \cite{Dain-Reiris11}, using minimal stable surfaces. It is remarkable that since this result is quasi-local, there is no asymptotically flatness requirement concerning the initial data. We only need the existence of stable minimal surfaces located at the axis, with aligned angular momenta (in order to maintain axial symmetry). It is clear that in order for these surfaces to have a non vanishing Komar angular momentum a non trivial topology is needed. One way to think about this is imagining that each surface encloses a puncture. 

A surface $\Sigma$ is called minimal if its mean curvature $\chi$ vanishes, and it is called stable if it is a local minimum of the area functional \eqref{areaf}. This can be precisely stated in the following form. Consider a flux of surfaces $F_t:\mathbb R\times S^2\to \Sigma$ parametrized by $t\in \mathbb R$ such that $F|_{t=0}(S^2)=\Sigma$ and $\dot F^a|_{t=0}=\alpha n^a$ where dot denotes derivative with respect to $t$, $n^a$ is the unit normal to $\Sigma$ and $\alpha$ is in principle an arbitrary function on $\Sigma$. Then the stability condition can be writen as
\begin{equation}\label{cond}
 \ddot A|_{t=0}=\int\alpha\dot\chi dS_\gamma,
\end{equation}
where
\begin{equation}\label{dotchi}
 \dot\chi=-\Delta_\gamma\alpha-(R-R_\gamma+\chi_{ab}\chi^{ab})\alpha,
\end{equation}
$\chi_{ab}$ is the second fundamental form of $\Sigma$, $R$ is the Ricci scalar associated to the 3-metric $g_{ab}$ (see equation \eqref{3metric}) and $\Delta_\gamma$ and $R_\gamma$ are the Laplace operator and the Ricci scalar associated with the 2-metric $\gamma_{ab}$.

As in the previous section, we allow the function $q$ in the 3-metric to have non zero values at the bounded components $\Gamma_i$. This leads us to the following result, which is an extension of theorem 1 in \cite{Dain-Reiris11}.

\begin{theorem}\label{teoarea}
Consider an axisymmetric, vacuum and maximal initial data, with a non negative cosmological constant and $N\geq1$ orientable closed stable minimal axially symmetric surfaces $\Sigma_{i}$. Then
\begin{equation}\label{ine2}
\sqrt{1+4\mathcal F_{i-1}}\sqrt{1+4\mathcal F_{i}}\geq \frac{8\pi |J_i|}{A_i}
\qquad i=1,..N
\end{equation}
where $J_i$ and $A_i$ are the angular momentum and area of $\Sigma_i$ respectively. 
\end{theorem}

Before giving the proof of theorem \ref{teoarea}, we want to remark that the equality in \eqref{ine2} is achieved by a solution describing two counter-rotating identical extreme Kerr black holes separated by a massless strut given by Manko \textit{et al}, \cite{Manko08}. The explicit values for the separation distance, ratio $A/8\pi J$ and the force in the extreme case as
\begin{equation}
 L=2m\frac{a^2+m^2}{a^2-m^2}, \qquad \frac{A}{8\pi |J|}=\frac{2|J|}{a^ 2+m^ 2},\qquad \mathcal F=\frac{(a^ 2-m^ 2)^ 2}{16J^ 2}
\end{equation}
where $J=am$ and $J$ is the angular momentum of each black hole, and $m$ its mass parameter (also, $2m$ is the total ADM mass). We note that $a^2>m^2$ holds for the extreme counter-rotating constituents, unlike in the case of a single Kerr black hole, for which $a^2=m^2$. It is also clear from the above expressions that as $m\to a^-$, then the separation distance becomes arbitrarily large, and the mutual force becomes arbitrarily small.

In case that the force is positive, which is the expected outcome, the quantity $A/8\pi |J|$ can be less than unity, a situation that can not happen for regular black holes. Nevertheless, a type of variational characterization of extreme Kerr throat is still valid in the presence of a conical singularity. This is due to the fact that in the variational principle used to derive the above inequality (see \cite{Acena11}), the function $q$ plays no rol, the relevant functions are the conformal factor $e^ \sigma$ and the twist potential $\omega$.

The inequality \eqref{aj1} has been recently extended in \cite{GabachJaramillo11}, \cite{GabachJaramilloReiris11} to include electric and magnetic charges and to apply also to stable marginally outermost trapped surfaces satisfying the dominant energy condition.To be more precise, consider a surface $\Sigma$ embedded in a spacetime, with null normal vectors $\ell^ a$ and $k^ a$ such that $\ell^ ak_a=-1$ and $\ell^ a$ is outward pointing. Let $\theta^ {(\ell)}$ be the expansion associated with the null normal $\ell^ a$. Then, $\Sigma$ is a stable marginally outermost trapped surface if $\theta^ {(\ell)}=0$ and if there exists an outgoing vector $X^ a=\gamma\ell^ a-\psi k^ a$ with $\gamma\geq0$ and $\psi>0$ such that $\delta_X\theta^ {(\ell)}\geq0$. Here $\delta_X$ denotes the deformation operator on $\Sigma$ that controls the infinitesimal variations of geometric objects defined on $\Sigma$ under an infiniteimal deformation of the surface along the vector $X^ a$. By following the same lines as in theorem \eqref{teoarea} based on the proof of \cite{GabachJaramillo11}, \cite{GabachJaramilloReiris11} one obtains the following corollary
\begin{corollary}\label{teoareaq}
Consider an axisymmetric, stable marginally outermost trapped surface $\Sigma_{i}$, with a non negative cosmological constant, satisfying the dominant energy condition. Then
\begin{equation}
\sqrt{1+4\mathcal F_{i-1}}\sqrt{1+4\mathcal F_{i}}\geq\frac{\sqrt{(8\pi J_i)^2+(4\pi Q_{iE}^2)^2+(4\pi Q_{iB}^2)^2}}{A_i} \qquad i=1,..N.
\end{equation}
where $J_i$, $Q_{iE}$, $Q_{iB}$ and $A_i$ are the angular momentum, electric charge, magnetic charge and area of $\Sigma_i$ respectively. $\mathcal F_i$ and $\mathcal F_{i-1}$ are the forces acting on $\Sigma_i$ along the axis componentes  $\Gamma_i$ and $\Gamma_{i-1}$.
\end{corollary}

In the case of zero angular momenta, the inequality is saturated for the Majumdar-Papapetrou solution with $J_i=Q_{iB}=0$, $A_i=4\pi Q_{iE}^2$ and vanishing forces. The same thing does not seem plausible for Kerr black holes, since numerical results \cite{DainOrtiz09} suggest  that a configuration with two Kerr black holes having $a=m$ leads to a  positive force. 

\vspace{0.5cm}
\noindent\textit{Proof of theorem \ref{teoarea}}. We follow the lines of \cite{Dain-Reiris11} and refer the reader to this article for more details. Choose a coordinate system such that the determinant of the induced metric $\gamma_{ab}$ on $\Sigma_i$ is 
\begin{equation}
 \sqrt{det(\gamma)}=e^{c_i}\sin\theta
\end{equation}
where $c_i$ is a constant. Then, the metric can be written as
\begin{equation}
 \gamma_{ab}=e^{\sigma_i}\left(e^{2q_i}(d\rho^2+dz^2)+\rho^2\sin^2\theta d\phi^2\right)
\end{equation}
with $\sigma_i+q_i=c_i$ and
\begin{equation}
 (\sigma_i+q_i)|_{\theta=0}=(\sigma_i+q_i)|_{\theta=\pi}=c_i
\end{equation}

Now multiply $\dot\chi$, given by \eqref{dotchi} by $\alpha$, use the constraint to express the Ricci scalar $R$ in terms of $K_{ab}$, integrate over $\Sigma_i$ and use condition $\chi=0$, the stability condition \eqref{cond} and the bound \eqref{boundk}. Then we find, disregarding the non negative terms
\begin{equation}\label{ec1}
 \int(|D\alpha|^2+\frac{1}{2}R_\gamma\alpha^2)e^cdS\geq\int\frac{1}{4}\frac{\omega_i'^2}{\eta^2}e^{\sigma_i-2c_i}\alpha^2e^{c_i}dS,
\end{equation}
where a prime denotes derivative with respect to $\theta$. 
Now, using that
\begin{equation}
 R_\gamma=\frac{e^{\sigma_i-2c_i}}{\sin\theta}(2q_i'\cos\theta+\sin\theta\sigma_i'q_i'+2\sin\theta-(\sin\theta\sigma_i')')
\end{equation}
and choosing
\begin{equation}
 \alpha=e^{c_i-\sigma_i/2}
\end{equation}
we obtain
\begin{equation}\label{ec2}
 4\pi(c_i+1)-2\pi(q_i(\pi)+q_i(0))-\int(\sigma_i+\frac{1}{4}\sigma_i'^2)dS\geq\int\frac{1}{4}\frac{\omega_i'^2}{\eta^2}dS
\end{equation}
which can be written as
\begin{equation}
 4\pi(c_i+1)-2\pi(q_i(\pi)+q_i(0))\geq\frac{\pi}{2}\mathcal M_i
\end{equation}
where $\mathcal M_i$ is the functional introduced by \cite{Acena11} and used also by \cite{Dain-Reiris11} and \cite{Jaramillo11},
\begin{equation}
 \mathcal M_i:=\frac{1}{2\pi}\int4\sigma_i+\sigma_i'^2+\frac{\omega_i'^2}{\eta^2}dS
\end{equation}
Note that in going from \eqref{ec1} to \eqref{ec2} is  where we have used the non regularity of the metric at the axis. 
Now, in \cite{Acena11} it has been proven that for each $\mathcal M_i$, the following bonds holds
\begin{equation}
\mathcal M_i\geq 8(\ln(2|J_i|)+1),
\end{equation}
where $J_i$ is the angular momentum of $\Sigma_i$. Using the fact that $A_i=4\pi e^{c_i}$ we find 
\begin{equation}\label{ineq}
A_i\geq 8\pi|J_i|e^{\frac{q_i(0)+q_i(\pi)}{2}}
\end{equation}
Finally, using the expression for $\mathcal F_i$ in terms of $q_i$ we complete the proof of the theorem. \begin{flushright}
$\Box$
\end{flushright}

\section*{Acknowledgments}
Part of this work was developed at the Erwing Schr\"odinger Institut, in Vienna in August 2011. It is a pleasure to thank the organizers of the Dinamics in General Relativity Programme for their kind invitation. Also, special thanks go to Helmut Friedrich, Martin Reiris and Sergio Dain for useful comments and discussions.


\begin{thebibliography}{10}
\bibitem{Acena11}Acena, A.,  Dain, S. and Gabach Clement, M.E. 
Class. Quant. Grav. 28 105014 (2011). gr-qc/1012.2413.

\bibitem{Bach21} Bach, R. and Weyl, H. . 
Math. Z., 13 132 (1921).



\bibitem{ChruscielGalloway10}
Chru\'sciel, P., Galloway, G. and Pollack, D. gr-qc/1004.1016, 2010.

\bibitem{Chrusciel08a} Chru\'sciel, P., (2008) 
Annals Phys. 323 2566-2590. gr-qc/0710.3680.

\bibitem{Chrusciel08} Chru\'sciel, P., Li, Y. and Weinstein, G. (2008) 
Annals Phys. 323 2591-2613. gr-qc/0712.4064.

\bibitem{Costa00}
Costa, M. and Perry, M. Nucl.Phys. B591 (2000) 469-487. gr-qc/0008106.

\bibitem{Dainreview11}  
Dain, S. gr-qc/1111.3615, 2011.

\bibitem{Dain-Reiris11}Dain, S. and Reiris, M.  
Phys. Rev. Lett. 107 051101 (2011). gr-qc/1102.5215. 

\bibitem{DainOrtiz09} Dain, S. and Ortiz, O. 
Phys. Rev. D 80 024045 (2009). gr-qc/0905.0708.

\bibitem{Dain06} Dain, S. 
J. Diff. Geom. 79 33-67 (2006). gr-qc/0606105.

\bibitem{Einstein36}
Einstein, A. and Rosen, N. Phys. Rev. 49 (1936) 404.

\bibitem{Gabach11}Gabach Clement, M.E. (2011) 
gr-qc/1102.3834.

\bibitem{GabachJaramillo11}
Gabach Clement, M.E. and Jaramillo, J.L. (2011) gr-qc/1111.6248 .

\bibitem{GabachJaramilloReiris11}
Gabach Clement, M.E., Jaramillo, J.L.  and Reiris, M. (2011) Work in preparation.

\bibitem{Herdeiro09}
Herdeiro, C.,  Kleihaus, B., Kunz, J. and Radu, E. Phys.Rev. D81 (2010) 064013. gr-qc/09123386.

\bibitem{Neugebauer11b} Hennig, J. and Neugebauer, G. (2011), 
gr-qc/1103.5248.

\bibitem{Jaramillo11}Jaramillo, J.L., Reiris, M. and Dain, S. (2011), 
gr-qc/1106.3743.

\bibitem{Li91} Li, Y. and Tian, G. 
Manuscripta Math. 73 83-89 (1991).

\bibitem{Li92} Li, Y. and Tian, G.  
Comm. Math. Phys. 149 1-30 (1992).

\bibitem{Costa09} Lopes Costa, J. (2009) 
gr-qc/0912.0838.





\bibitem{Manko08}Manko, V.S., Rodchenko, E.D., Ruiz, E. and Sadovnikov, B.I.  
Phys. Rev. D 78 124014 (2088). gr-qc/0809.2422.  

\bibitem{Neugebauer11} Neugebauer, G. and Hennig, J. (2011), 
gr-qc/1105.5830.



\bibitem{Neugebauer09} Neugebauer, G. and Hennig, J. , 
Gen. Rel. Grav. 41 2113-2130 (2010). gr-qc/905.4179.

\bibitem{Varzugin97} Varzugin, G.G. (1997), 
Theoretical and Mathematical Physics 111 3 1997.

\bibitem{Varzugin98} Varzugin, G.  
Theor. Math. Phys. 116 1024-1033 (1998).

\bibitem{Weinstein92} Weinstein, G.
Communications on Pure and Applied Mathematics, 45: 1183-1203 (1992).

\bibitem{Weinstein94} Weinstein, G.  
Transactions of the American Mathematical Society, 343 2 899 (1994).


\bibitem{Weinstein90} Weinstein, G.  
Communications on Pure and Applied Mathematics, 43: 903-948 (1990).
\end{thebibliography}
\end{document}